\documentstyle[12pt]{article}
\begin{document}
\title{\bf On the possible space-time fractality of the emitting source}
\author{O.V.Utyuzh$^{1}$\thanks{e-mail: utyuzh@fuw.edu.pl},~
G.Wilk$^{1}$\thanks{e-mail: wilk@fuw.edu.pl} and Z.W\l odarczyk$^{2}$
\thanks{e-mail: wlod@pu.kielce.pl}\\[2ex] 
$^1${\it The Andrzej So\l tan Institute for Nuclear Studies}\\
    {\it Ho\.za 69; 00-689 Warsaw, Poland}\\
$^2${\it Institute of Physics, Pedagogical University}\\
    {\it  Konopnickiej 15; 25-405 Kielce, Poland}}  
\date{\today}
\maketitle

\begin{abstract}
Using simple space-time implementation of the random cascade model we 
investigate numerically a conjecture made some time ago which was
joining the intermittent behaviour of spectra of emitted particles
with the possible fractal structure of the emitting source. We
demonstrate that such details are seen, as expected, in the
Bose-Einstein correlations between identical particles. \\ 


\end{abstract}

\newpage

\section{Introduction}

The multiparticle spectra of secondaries produced in high energy
collision processes are the most abundant sources of our knowledge of
the dynamics of such processes. Among others, two features emerging
from the analysis of these spectra are of particular interest: 
$(i)$ the so called intermittent behaviour observed in many
experiments in the analysis of factorial moments of spectra of
produced secondaries and $(ii)$ the Bose-Einstein correlations (BEC)
observed between identical particles. Whereas the former seems to
indicate the existence of some (multi) fractal structure of the
production process \cite{CAS} the latter are established as, by now 
the most important source of our knowledge on the space-time
aspects of the multiparticle production processes \cite{BEC}.\\

Some time ago it was argued \cite{B,B98} that, in order to make
both effects compatible with each other, the emitting source should
fluctuate in size in a scale-invariant (i.e., power-like) way. This
can be achieved in two ways: $(i)$ either the shape of the interaction
region is regular but its size fluctuates from event to event
according to some power-like scaling law $(ii)$ or the interaction
region itself is a self-similar fractal extending over a very large
volume \cite{B,PPT}.\\ 

In this work we would like to investigate in more detail to what
extent the BEC is sensitive to the possible space-time fractality of
the emission source. To this end we shall use a simple self-similar
cascade process \cite{SS} in which the final particles are produced in the
sequential two-body decays of some original mass $M$. For our purpose
we shall extend it by introducing the simple (classical) space-time
development of the cascade and by adding the kind of BEC
``afterburner'' along the lines advocated recently in \cite{GEHW}. \\

It is widely expected that every cascade model has automatically
built in the intermittent behaviour of spectra of observed particles
\cite{ZAL}. Although this statement is true and obvious for the
models based on random multiplicative processes in some chosen
observed variables (like energy, rapidity or azimuthal angle) it is
highly non trivial in the case of cascades (or multiplicative
processes) proceeding in variable(s) not directly mesurable but
nevertheless of great dynamical importance (as, for  example, masses
of some intermediate objects occuring during the production process
\cite{SS}). In the purely mathematical case, where cascade process
proceeds {\it ad infinitum}, one eventually arrives at some
space-time fractal picture of the production process. However,
both the finite masses of produced secondaries and limited energy
(or mass $M$) stored in the emitting source prevent the full and
distinct development of such a fractal structure \cite{CARR}. One 
must therefore be satisfied with only some limited and mostly indirect
presence or signals of such structure. If established it would,
however, be important for our knowledge of the dynamics of the
multiparticle production process.\\

Such a fractal structure in phase-space can generate a
similar structure in the space-time picture of the hadronization
process. Our aim here is to demonstrate to what extent they influence
BEC. In the next two Sections we shall then provide, respectively,
the phase-space and space-time characteristics of a simple cascade
model used for that purpose. Section 4 contains our main results
showing the BEC features emerging from our model. Section 4 contains
a summary of our results and conclusions.\\ 

\section{Phase-space characteristic of the cascade model used}

We shall model the emitting source of mass $M$ by the usual $(1
\rightarrow 2$) random cascade process employed already in \cite{SS},
$M\, \longrightarrow \, M_1\, +\, M_2$, in which the initial mass $M$
``decays'' into two masses $M_{1,2} = k_{1,2}\cdot M$ in such a way
that $k_1 + k_2 < 1$, i.e., a part of $M$ equal to $(1- k_1 - k_2) M$
is transformed to kinetic energies of  the decay products $M_{1,2}$.
The process repeats itself (see Fig. 1) 
until $M_{1,2} \ge \mu$ ($\mu$ being the mass of the produced
particles)
with successive branchings occurring sequentially and independently
from each other, and with different values of $k_{1,2}$ at each
branching, but with energy-momentum conservation imposed at
each step. For different choices of dimensionality $D$ of our cascade
process (provided by the restrictions for the possible directions of
flights of the decay products in each vertex) be it $D=1$ or $D=3$
dimensional (isotropic) and for different choices of the decay
parameters $k_{1,2}$ at each vertex, we are essentially covering
an enormously vast variety of different possible production schemes
ranging from the  essentially one-dimensional strings to thermal-like
fireballs.\\ 

Of special interest is the case of a one-dimensional cascade for which
one can provide analytic formulae for the rapidities $Y_{1,2}$ of the
decay product at each vertex given in the rest frame of the parent
mass in this vertex. They depend solely on the decay parameters at
this vertex, $k_{1,2}$:    
\begin{eqnarray}
Y_1\, &=&\, \pm \ln \left[ \frac{1}{2k_1} \left(1\, +\, k_1^2\, -\, k_2^2\right)\,
              +\, \frac{1}{2k_1}\, \sqrt{\Delta} \right] , \nonumber\\
Y_2\, &=&\, \mp \ln \left[ \frac{1}{2k_2 }\left(1\, -\, k_1^2\, +\, k_2^2\right)\,
              +\, \frac{1}{2k_2}\, \sqrt{\Delta} \right] , \label{eq:YY}\\
       && {\rm where}\qquad 
       \Delta\, =\, \left( 1 - k_1^2 + k_2^2 \right)^2 - 4 k_2^2 .   \nonumber
\end{eqnarray}                          
      
Two limiting cases can be distinguished here: $(i)$ - totally
symmetric and $(ii)$ maximaly asymmetric cascades. In the case of a
totally symmetric cascade decay parameters are equal and the same for
all vertices, $k_{1,2}=k$. In this case the finally produced particles
occur only at the very end of the cascade process and the amount of
energy allocated to the production is maximal. Because the number of
possible branchings characterizing the length of the cascade is equal
to $L_{max} = \ln \frac{M}{\mu} / \ln\frac{1}{k}$ (where $\mu =
\sqrt{m^2_0 + \langle p_T\rangle^2}$), the multiplicity of produced
secondaries is given by the following formula: 
\begin{equation}
N_{s}\, =\, 2^{L_{max}}\, =\, 
           \left( \frac{M}{\mu}\right)^{d_F},
           \qquad d_F\, =\, \frac{\ln 2}{\ln\frac{1}{k}} .
            \label{eq:NSYM}
\end{equation}
According to \cite{SS,ID} the exponent $d_F$ is formaly nothing but
a generalized (fractal) dimension of the fractal structure of phase
space formed by our cascade. The utility of such notion is, however,
greatly reduced because of the necessary limited length of our
cascades \cite{CARR}. Notice a kind of scaling in (\ref{eq:NSYM})
where, for a fixed ratio $\frac{M}{\mu}$ the observed multiplicity
$N_s$ depends solely on the decay parameter $k$. The characteristic
power-like behaviour of $N_{s}(M)$ in (\ref{eq:NSYM}) is normally
atributed to thermal models. For example, for $k=\frac{1}{4}$ one has
$N_{s} \sim M^{\frac{1}{2}}$, which in thermal models would
correspond to the ideal gas equation of state with velocity of sound
$c_0 = \frac{1}{\sqrt{3}}$ \cite{HYDRO}. The same behaviour (on
average) is obtained for $k_{1,2}$ chosen randomly from a triangle
distribution $P(k) = (1 - k)^a$ with $a\simeq 1$ which will be
therefore used in all our numerical calculations. The fact that $N_s
\in \left(2,\frac{M}{\mu}\right)$ means that decay parameters are
limited to $k \in (\frac{\mu}{M}, 1)$.\\

For maximally asymmetric cascades $k_1 = \frac{\mu}{M}$ and $k_2= k$,
i.e., at each step one has always a single final particle of
transverse mass $\mu$ produced against some recoil mass $M_1 =
kM$: $M \rightarrow \mu + M_1$. The amount of kinetic energy
allocated to the produced secondaries is now maximal. The
corresponding rapidities $Y_1$ and $Y_2$ are now given by eqs.
(\ref{eq:YY}) with, respectively, $k_1=\frac{\mu}{M}$ and $k_2 = k$.
Also here the resultant multiplicity $N_{a}$ is given by the length 
$L_{max}$ of the cascade, which is limited by the condition $M_l =
k^l\, M \geq \mu $. It means therefore that $L_{max} =\ln
\frac{M}{\mu}/\ln \frac{1}{k}$  and the corresponding multiplicity is
\begin{equation}
N_{a}\, =\, 1\, +\, L_{max} \, =\, 
     1\, +\, \frac{1}{\ln\frac{1}{k}}\cdot
                      \ln \frac{M}{\mu} . \label{eq:NASYM}
\end{equation}                      
The energy dependence of $N_a$ is now logarithmic (in thermal models
it would correspond to a one dimensional fireball with $c_0 \rightarrow 1$
\cite{1DIM}) but the same kind of scaling as in eq. (\ref{eq:NSYM})
is present also here.\\

Our simple model therefore covers all possible energy dependencies of
the multiplicities of produced particles which depend solely on decay
parameters $k_{1,2}$ of the cascade and scale in the ratio
$\frac{M}{\mu}$. This remains true for both one and three dimensional
cascades.\\ 

In Fig. 2 one can see examples of rapidity distributions
$\frac{1}{N}\frac{dN}{dy}$ calculated for symmetric and asymmetric
$D=1$ cascades discussed above. They are compared there with most
probable distributions in one-dimension obtained by means of 
information theory arguments \cite{EIWW}  
\begin{equation}
f_{IT}(y)\, =\, \frac{1}{Z}\cdot \exp\left[ - \beta\cdot\mu \cosh y\right]
\label{eq:INFO},
\end{equation}
where $\int dy f(y) = 1$ (what defines $Z$) and $\beta = \beta(M,N)$ is
the corresponding lagrange multiplier ensuring proper conservation of
energy-momentum in the case when $N$ particles, each of transverse mass
$\mu$, are produced from the source of mass $M$. Contrary to the case
of production via the cascade process, nothing is now said about the
production mechanism. It is just tacidly assumed that all produced
particles occur, in a sense, instantenously in the whole allowed 
phase space with the weights provided by eq. (\ref{eq:INFO}), which
was obtained by the maximalization of the suitably defined
information entropy corresponding to the production process under 
consideration \cite{EIWW}. It occurs that for a wide range of $M$ and
$N$ the quantity $\bar{\beta} = \beta\frac{M}{N}$ is (almost)
constant as a function of energy per particle $\bar{m} =
\frac{M}{N}$, i.e., also here one encounters a kind of scaling,
namely that $\frac{1}{N}\frac{dN}{dy} \sim F\left(z = \frac{\mu\,
\cosh y}{\bar{m}}\right)$.\\

The shape of the multiplicity distribution, $P(N)$, in our case of
a source with fixed ratio $\frac{M}{\mu}$ is given by distribution
$P(k_{1,2})$ of decay parameters $k_{1,2}$. We shall use a simple
triangle form for it (as already mentioned above) $P(k) = (1 - k)^a$, 
which for $a\simeq 1$ provides the commonly accepted energy behaviour
of the mean multiplicities $N(M) \sim M^{0.4\div0.5}$ as discussed
above. The example of $P(N)$ for $D=1$ cascades are shown in Fig. 3
($P(N)$ for $D=3$ cascades are the same). They exemplify three
different choices of the ratio $\frac{M}{\mu}$ ($\frac{10}{0.3} =
33.3,~\frac{40}{0.3} = 133.3$ and $\frac{100}{0.3} = 333.3$,
respectively). \\     

In Fig. 4, we show the behaviour of scaled (``horizontal'' \cite{BL})
factorial moments $F_l$ (for $l=2,3$) calculated for a one-dimensional
cascade in rapidity space as a function of number of bins $n_{bin} =
Y/\delta y$ (where $Y$ is taken as the corresponding  rapidity range
for the corresponding mass $M$ and $\delta y$ denotes the bin size
considered):  
\begin{equation}
F_l\, =\, n_{bin}^{l-1}\left\langle \, \sum^{n_{bin}}_{i=1}\,
             \frac{ n_i\left(n_i - 1    \right) \cdots
                       \left(n_i - l + 1\right)}
             {\langle N\rangle^l}\, \right\rangle  . \label{eq:Fh}
\end{equation}
The $N$ produced particles are distributed among $n_{bin}$ bins with
$n_i$ particles in the $i^{th}$ bin, i.e., $\sum^{n_{bin}}_{i=1}n_i =
N$.  The average is over all events.
It is interesting to note that our results, although obtained for
essentially the same type of cascade as discussed in \cite{SS},
apparently demonstrate much stronger intermittency signal than the
experimental one shown there. However, no fit to the data was attempted
in our case as we are concerned with the properties of a single
elementary source only (leaving the problem of their distribution in
mass $P(\frac{M}{\mu})$ aside). On the other hand, moments $F_l$ in
\cite{SS} were in reality not calculated but deduced from
experimental data by means of some simple formula obtained from
general (mathematical) fractal analysis of symmetric cascade
processes. The aim was to deduce from them the fractal dimensions
$d_F = \ln 2/\ln \frac{1}{k}$ of cascade process considered. As a
result the corresponding decay parameter in \cite{SS} turns out to be
very large, $k\simeq 0.45$, leading to $\langle N(M)\rangle \simeq
M^{0.8 \div 0.9}$ instead of expected $\langle N(M)\rangle \simeq
M^{0.4 \div 0.5}$ as discussed above \cite{HYDRO}.\\

\section{Space-time characteristic of the cascade model used}

We shall now endow our cascade in phase-space with space-time
elements (not addressed in  \cite{SS}). To this aim we introduce some
fictitious finite ``life time'' $t$ for each vertex mass $M_l$, which 
is allowed to fluctuate according to some prescribed distribution law
$\Gamma(t)$. This procedure is a purely classical one, i.e., we are
not treating $M_l$ as resonances, as was done, for example, in
\cite{PIS} on another occasion. Instead, they are regarded to be real
particles with masses given by the corresponding values of decay
parameters $k_{1,2}$ and with the respective velocities equal to
$\vec{\beta} = \frac{\vec{P}_{1,2}}{E_{1,2}}$
($(E_{1,2};\vec{P}_{1,2})$ are the energy-momenta of the
corresponding decay product given in the rest frame of the parent
mass in each vertex). The energy-momentum and charges are strictly
conserved in each vertex separately (this is another difference with
the information theory approach \cite{EIWW} where such conservation
laws are imposed on the whole process instead). As for a
decay/branching law we shall choose it in the simplest possible
exponential form: 
\begin{equation}
 \Gamma (t) \, =\, \frac{1}{\tau}\cdot
                   \exp{\left[ - \frac{t}{\tau}\right]}
               \label{eq:Gamma}
\end{equation}

It is straightforward to get our cascade model in the form of a
Monte Carlo code. The main features of a one-dimensional case has 
already been demonstrated above. The only difference between one and
three-dimensional cascades is in the fact that, whereas in the former
decay products can flow only along one, chosen direction, in the
latter in each vertex the flow direction is chosen randomly from the
isotropic angular distribution. To allow for some nonzero transverse
momentum in the one-dimensional case we are using the transverse mass
$\mu = 0.3$ GeV. For the three-dimensional cascade 
$\mu$ is instead set simply to the pion mass, $\mu = 0.14$ GeV. In
every case all decays are described in the rest frame of the
corresponding parent mass in a given vertex. To get the final
distributions, one has to perform a number of Lorentz transformations
to the rest frame of the initial source mass $M$. As output we are
getting in each run (event) a number $N_j$ of secondaries of mass
$\mu$ with both defined energy-momenta $\left[ E_j = \sqrt{\mu^2 +
\vec{P}^2_j}; \vec{P}_j\right]_{i=1, \dots, N_j}$ and space-time
coordinates $\left[t_j; \vec{r}_j\right]_{i=1, \dots, N_j}$ of the
last branching (i.e., the coordinates of birth of each particle).\\   

Fig. 5 shows densities $\rho(r)$ of points of production for all 
cases investigated here: for $D=1$ and $3$ dimensional cascades with
both constant and mass dependent evolution parameter $\tau$ and for
three choices of the source mass, $M=10,~40$ and $100$ GeV. As one
can see the widely expected (cf. \cite{B,PPT}) power-like behaviour
of cascading source 
\begin{equation}
\rho(r)\, \sim\,  \left(\frac{1}{r}\right)^L 
\qquad r > r_0 , \label{eq:scaling}
\end{equation}
is seen only (if at all) for $r > r_0$, i.e., for radii larger then
some (not sharply defined) radius $r_0$, value of which depends on
all parameters present here: mass $M$ of the source, dimensionality
$D$ and evolution parameter $\tau$ of the cascade. Below $r_0$ the
$\rho(r)$ is considerably bended, remaining even almost flat for 
$D=1$ cascades. Only for rapidly developing cascades (i.e., for $\tau
\sim \frac{1}{M}$) in $D=3$ dimensions, the expected scaling sets in
almost from the very beginning and it practically does not depend 
on the mass of the source. For the limiting case of $M=100$ GeV the
corresponding values of parameter $L$ vary from $L=1.89$ and $L=1.86$
for $\tau = 0.2$ and $1/M$ for one dimensional cascades to $L=2.78$
and $L=2.8$ for three dimensional cascades.\\

The shapes of $\rho(r)$ scale in the ratio $\left( \frac{M}{\mu}
\right)$ in the same way as the multiplicity distributions $P(N)$
discussed before. 
As desidered power-like behaviour \cite{B}-\cite{PPT} sets in
(at least approximately) only for long cascades (large values of
$\left( \frac{M}{\mu} \right)$) and/or for fast ones (small values of
$\tau$), it remains therefore to be checked whether (and to what
extent) such conditions are indeed met in the usual hadronic
processes. This point is, however, outside of the scope of the
present work.\\ 

\section{Cascades and BEC}

Let us proceed now to our main point, namely to the question of
whether one can see in BEC some special features which could be
attributed solely to the the branchings and to their space-time and
momentum space structure. At first glance, the answer seems to be
plainly negative as it is easy to check that the function  
\begin{equation}
C_2(Q=|p_i - p_j| )\, =\, 
             \frac{d N(p_i,p_j)}{dN(p_i)\, dN(p_j)} \label{eq:C2}
\end{equation}
does not show any structure of BEC type. It is also true if we endow
our cascade process with the production of charges of the type: 
$\left\{0\right\}  \rightarrow  \left\{+\right\}+\left\{-\right\}$, 
$\left\{+\right\}  \rightarrow  \left\{+\right\}+\left\{0\right\}$
and 
$\left\{-\right\}  \rightarrow  \left\{0\right\}+\left\{-\right\}$.
In this case $C_2$ calculated for like charge pairs also does not
show any correlations. That is, however, to be expected, because the
only way to have (\ref{eq:C2}) showing ``primordial'' BEC is to
introduce them from the very beginning, for example in the way done
recently in \cite{OMT}. Using the same information theory approach 
as in \cite{EIWW} but adding a new piece of information, namely that
the produced particles are mostly bosons and as such they should be
grouped together as much as possible in the phase space cells, the
authors of \cite{OMT} indeed obtained a substantial BEC signal, namely
$C_2 > 1$ for $p_i \rightarrow p_j$ in (\ref{eq:C2}). No space-time
structure was discussed in \cite{OMT}, however.\\ 

We cannot follow this prescription here without invoking a kind of
extremely difficult to formulate (or calculate) special final
state interactions between produced secondaries. Our cascade is
supposed to mimic the production process in its development, whereas
the information theory procedure of \cite{OMT} makes no statements
whatsoever about the development of the proces as such. It only 
provides the least biased and at the same time most probable
distributions, limited only by imposed constraints of the
energy-momentum and (mean) number of particles conservation
(reflection of which are the two lagrange multipliers,
$\beta(M,N)$ and $\mu(M,N)$, representing in terminology of the usual
thermal models the ``inverse temperature'' and ``chemical potential'',
respectively. We could, in principle, use the pseudopotential method
as, for example, advocated long ago in \cite{Z}, but this causes 
changes in the particle distributions and/or destroys the
energy-momentum balance which has to be later restored in a more or
less {\it ad hoc} way and it does not use the information on the
space-time structure of our results.\\

An open question is the possible existence of phase factors in every
branch point, which would endow each particular branch and through it
also the finally produced secondaries with some specific, possibly
path dependent phases. This question is, however, left open here and
it is understood that they are all set equal to unity. They would
be important to BEC correlations influencing especially values of
$C_2(Q=0)$, i.e., the so called degree of coherence/chaoticity
$\lambda$. We shall return to this problem elsewhere.\\

Because our aim is not data fitting but checking if, and to what
extent, the BEC via its $C_2$ observable is sensitive to different
choices of the cascade processes provided by different sets of
parameters, we have decided to use the ideas of the BEC
``afterburners'' advocated recently in \cite{GEHW}. And because we
are not so much interested in  particular values of the ``radius''
and ``coherence'' parameters $R$ and $\lambda$, but in the
systematics emerging from our study, we shall use for this
exploratory research the most primitive, classical version of such
``afterburner''. The procedure we use is therefore very simple. After
generating a set of $i=1,\dots,N_l$ particles for the $l^{th}$ event
we choose all pairs of the same sign and endow them with the weight
factors of the form  
\begin{equation}
C = 1 + \cos\left[ \left(r_i - r_j)(p_i - p_j\right)\right] \label{eq:C}
\end{equation}
where $r_i = (t_i,\vec{r}_i)$ and $p_i = (E_i,\vec{p}_i)$ for a given
particle.\\

The results obtained from $N_{event} = 50000$ events are presented in
Fig. 6. They are displayed for the same sequence of parameters $M$
(mass of the source), $D$ (its dimensionality) and $\tau$ (the
evolution parameter) as in Fig. 5. The characteristic feature to be
noted is a substantial difference between $D=1$ and $D=3$ dimensional
cascades both in the widths of the $C_2(Q)$ and their shapes. Whereas
the former are more exponential-like the latter are more gaussian-like
with a noticeably tendence to flattening out at very small values of
$Q$. Also values of intercepts, $C_2(Q=0)$, are noticeable lower for
$D=3$ cascades. There is also a difference between ``slow'' (constant
$\tau$) and ``fast'' ($\tau \sim \frac{1}{M}$) cascades, especially for
$D=1$ ones. The former lead to substantially different shapes in this
case. In $D=3$ this effect is not so visible, although it is also
present. The length of the cascade (i.e., the radius of the
production region, cf. discussion of density $\rho$ before) dictates
the width of $C_2(Q)$. However, the $\left(\frac{M}{\mu}\right)$
scaling observed before in multiplicity distributions and in shapes
of source functions is lost here. This is because $C_2$ depends on
the differences of the momenta $p=\mu\, \cosh y$, which do not scale
in $\frac{M}{\mu}$. The flattening mentioned above together with
$C_2(0) < 2$ for $D=3$ cascades are the most distinctive signature of
the fractal structure combined with $D=3$ dimensionality of the
cascade. The correlations of the position-momentum type existing here
as in all flow phenomena are, in the case of $D=3$ cascades, not
necessarily vanishing for very small differences in positions or
momenta between particles under consideration. The reason is that our
space-time structure of the process can have in $D=3$ a kind of
``holes'', i.e., regions in which the number of produced particles is
very small. This is perhaps the most characteristic observation for
fractal (i.e., cascade) processes of the type considered here.\\ 

\section{Summary and conclusions}

In this work we have addressed the problem of the possible space-time
fractal structure of the hadronic production process. It is
complementary to the possible (multi) fractality leading to 
fluctuations in the multiparticle distributions and to the possible
fractality claimed to exist already on the level of hadronic
structure \cite{DL}. Although there is a vast literature concerning
the possible (multi)fractality in momentum space \cite{HWA}
its space-time aspects are not yet fully recognized with
\cite{B}-\cite{PPT} remaining so far the only representative
investigations in this field. Our aim was to extent this
investigation a bit further by essentially repeating ideas proposed 
in \cite{B}-\cite{PPT} in a numerical form that allowed us to check
in more detail the conjectures made there, showing the limits of their
applicability.\\ 

Our simple model posseses all features called for in \cite{B} -
\cite{PPT}: it shows both intermittency in the phase space
(demonstrated in the limiting case of one-dimensional cascade
explicitely in Fig. 4) and (approximate) power law distribution of
the production points in the space-time (cf. Fig. 5). As we have more
constraints on the phase space behaviour imposed by the, for example,
expected energy dependence of the multiplicity $\langle N\rangle$, we
have not much freedom in choosing decay parameters $k$. 
Our distribution $P(k)$ is surely not the only possible one. 
However,
whatever shape we choose for $P(k)$ it should reproduce the expected
energy dependence of multiplicity, $\langle N(M)\rangle \sim
M^{0.4\div0.5}$. Therefore the only really free parameters in our
study were the time evolution parameter $\tau$ and dimensionality of
the cascade, $D$. Here only two extreme values of $D=1$ and $D=3$
were studied and two also, in a sense, extreme behaviours of $\tau
=const$ and $\tau \sim \frac{1}{M}$ were used. All three: $M,~D$ and
$\tau$ were found to influence the $C_2(Q)$ observable characterising
BEC, cf. Fig. 6. \\ 

We therefore conclude that BEC are, indeed, substantially influenced
by the fact that our process is of the cascade type as was
anticipated in \cite{B}-\cite{PPT}, although probably not to the
extent expected there (which, however, has not been quantified
there). However, in practical applications, i.e., in the eventual
fitting of experimental data, there are many points which need
further clarification. The most important is the fact that data are
usually collected for a range of masses $M$ and among directly
produced particles are also resonances. Therefore, one has first to
specify the form of the distribution $P(\frac{M}{\mu})$ which will
influence to some extent our results. In particular changes in $\mu$
due to the production from resonances will shorten our cascade
considerably. The possible effect can be to some extent deduced from
our results by comparing data with $M=100$ GeV with those with $M=40$
GeV and $M=10$ GeV. The BEC will be effective only in conjunction with
precise studies of distributions in the phase-space (like
$P(\frac{M}{\mu})$, intermittency and momentum and rapidity
distributions. These studies should to some extent fix the distribution
of decay parameters $P(k_{i,j})$. 
Only then one can fit data to different parameters $\tau$ 
characterizing the space-time structure of the source.\\

One should realize at this point that there were already attempts to
study the BEC with power-like (Lorenzian type) shape of source
function: $\rho(\xi) = \frac{3}{4\pi^2R^4}\frac{1}{\left(1 +
\xi^2/R^2\right)^{5/2}}$ which leads to $C_2 = 1 + \exp(-2RQ)$ (with
$\xi^2 =  \left( x^2 + y^2 + z^2 \right) + (ct)^2 $ and $Q$ defined
as in our case) \cite{MB}. Such a form is nearest to our $\rho(r)$ and,
as it turns out, gives the best fit (in terms of the
$\chi^2$-values) to data considered in \cite{MB}. However, 
differences between this fit and other more conventional ones (i.e.,
based on gaussian or exponential shapes of the source) were not
dramatic. This means, that in reality it will be very difficult to
establish by means of BEC the possible existence of fractal structure
of the emitting source. Perhaps the event-by-event analysis of data
with some preselection of the initial conditions (in terms of energy,
centrality, multiplicity etc.) will be necessary in order to perform
such investigations. \\

Our approach must be regarded as preliminary because of our
choice of treatment of BEC. Notwithstanding its obvious deficiencies
(already mentioned in \cite{GEHW}) it seems, however, fully adequate
for the present study which is, as mentioned above, of only limited
scope. However, even in such form it seems to indicate the relevance
of the fact of a possible fractal structure of the space-time of the
emitting region preventing particles from different branches to be in
the same emitting cell irrespectively of the  smallness of
differences in their positions or momenta. This stresses the
problem of the distance in the cascade and the like, recently
discussed in \cite{MV}.

\newpage

\noindent
{\bf Figure Captions:}\\

\begin{itemize}

 \item[{\bf Fig. 1}] The scheme of our cascade process.

 \item[{\bf Fig. 2}] Example of rapidity distributions of secondaries 
                     for totally symmetric $(a)$ and totally
                     asymmetric one-dimensional cascades with asymmetric 
                     $(b)$ and symmetric $(c)$ emission of 
                     particles calculated for $M=40$ GeV, $\mu=0.3$.
                     Histograms are for fixed $k$: $P(k)=\delta(k-0.25)$.
                     Open symbols display results for cascades with 
                     $k_{1,2}$ distributed randomly according to 
                     $P(k) = (1 - k)^a$ with $a=1$ (in both cases
                     multiplicities are the same: $N_{(a)} \simeq 11.5$ and
                     $N_{(b,c)} \simeq 4.5$). Full lines present
                     the most probable (one dimensional) thermal-like 
                     distributions given by (\ref{eq:INFO}) 
                     with $\beta = 0.028$ for $(a)$ and $\beta =
                     -0.133$ for $(b)$ and $(c)$, respectively,
                     (calculated as in \cite{EIWW} for $N_{(a)} = 11.5$ and
                     $N_{(b,c)} = 4.5$).
                     
 \item[{\bf Fig. 3}] Multiplicity distributions $P(N)$ for
                     (one dimensional) cascades of masses 
                     $M=10,~40,~100$ GeV (for $\mu = 0.3$ GeV and 
                     $k_{1,2}$ given by the same triangle 
                     distributions $P(k)$ as in Fig. 2):
                     $(a)$ - symmetric case with respective mean values 
                     $\langle N\rangle =\, 7.39, ~14.67, ~22.76$ 
                     and dispersions $\sigma = \, 2.37, ~5.12, ~8.47$;
                     $(b)$ - asymmetric case with $\langle N\rangle 
                     =\, 3.97, 4.88, ~5.48$ and $\sigma =\, 1.07, 
                     ~1.27, ~1.41$.

 \item[{\bf Fig. 4}] Example of $2^{nd}$ and $3^{th}$ scaled 
                     ``horizontal'' moments $F_l$ as function 
                     of the number of bins $n_{bin} = Y/\delta y$
                     (where $Y$ is the rapidity range considered 
                     and $\delta y$ the bin size) for one-dimensional 
                     cascade of $M= 40$ GeV and $\mu =0.3$ GeV with 
                     $k_1=k_2=0.25$ and $k_{1,2}$ chosen randomly in 
                     the same way as in Fig. 3. The results for 
                     $M=100$ GeV are essentially identical.

 \item[{\bf Fig. 5}] Density distribution of the production points 
                     $\rho(r)$ for one-dimensional cascades 
                     ($r = \sqrt{x^2}$, $\mu = 0.3$ GeV) - left panels, 
                     and for three-dimensional cascades 
                     ($r= \sqrt{x^2+y^2+z^2}$, $\mu = 0.14$ GeV) - 
                     right panels. Two different choices of the
                     evolution parameter $\tau$ are considered: 
                     $\tau = 0.2$ fm - upper panels, and 
                     $\tau = 0.2/M$ (in fm, the mass $M$ is the parent 
                     mass in a given vertex) - lower panels. Each panel
                     shows results for three different masses $M$ of
                     the source: $M = 10,40$ and $100$ GeV. In all cases 
                     $k$ is chosen from the same triangle 
                     distribution distribution as in Fig. 2. 

 \item[{\bf Fig. 6}] The $C_2(Q=|p_i - p_j|)$ for the sources presented
                     in Fig. 5: left panels - one-dimensional cascades,
                     right panels - three-dimensional cascades;
                     upper panels - time evolution parameter is set 
                     constant and equal $\tau = 0.2$ fm, lower panels
                     - time evolution parameter is chosen as 
                     $\tau = 0.2/M$ (in fm, the mass $M$ is the parent 
                     mass in a given vertex). Each panel
                     shows results for three different masses $M$ of
                     the source: $M = 10,40$ and $100$ GeV. In all cases 
                     $k$ is chosen from the same triangle 
                     distribution distribution as in Fig. 2. 
                     
\end{itemize}

\newpage
\begin{figure}[h]
\setlength{\unitlength}{1cm}
\begin{picture}(25.,16.5)
\includegraphics{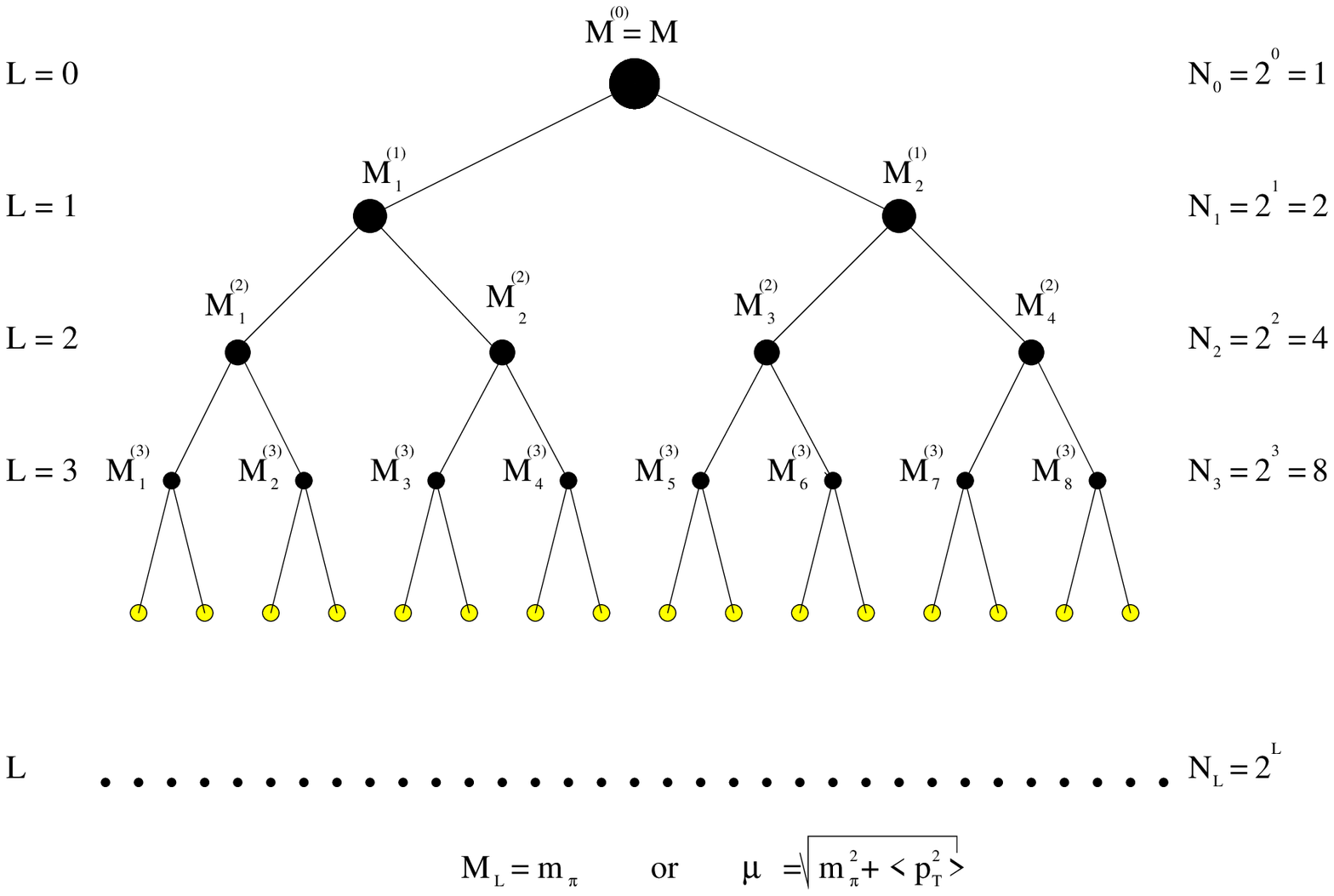}
\end{picture}
\end{figure}

\newpage
\begin{figure}[h]
\setlength{\unitlength}{1cm}
\begin{picture}(25.,16.5)
\includegraphics{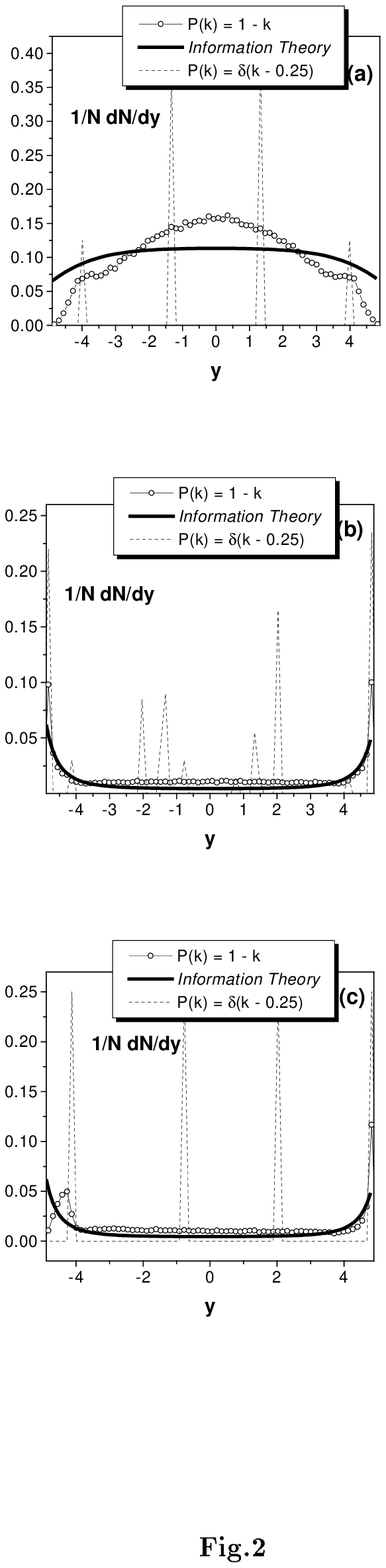}
\end{picture}
\end{figure}

\newpage
\begin{figure}[h]
\setlength{\unitlength}{1cm}
\begin{picture}(25.,16.5)
\includegraphics{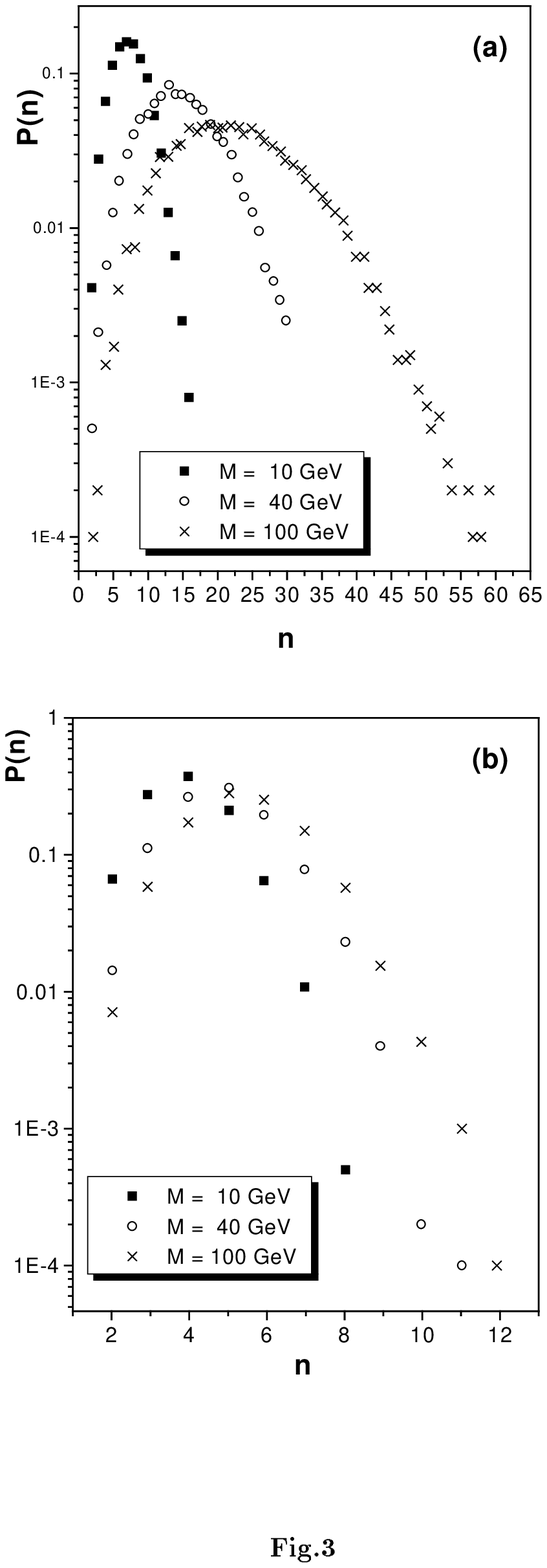}
\end{picture}
\end{figure}

\newpage
\begin{figure}[h]
\setlength{\unitlength}{1cm}
\begin{picture}(25.,16.5)
\includegraphics{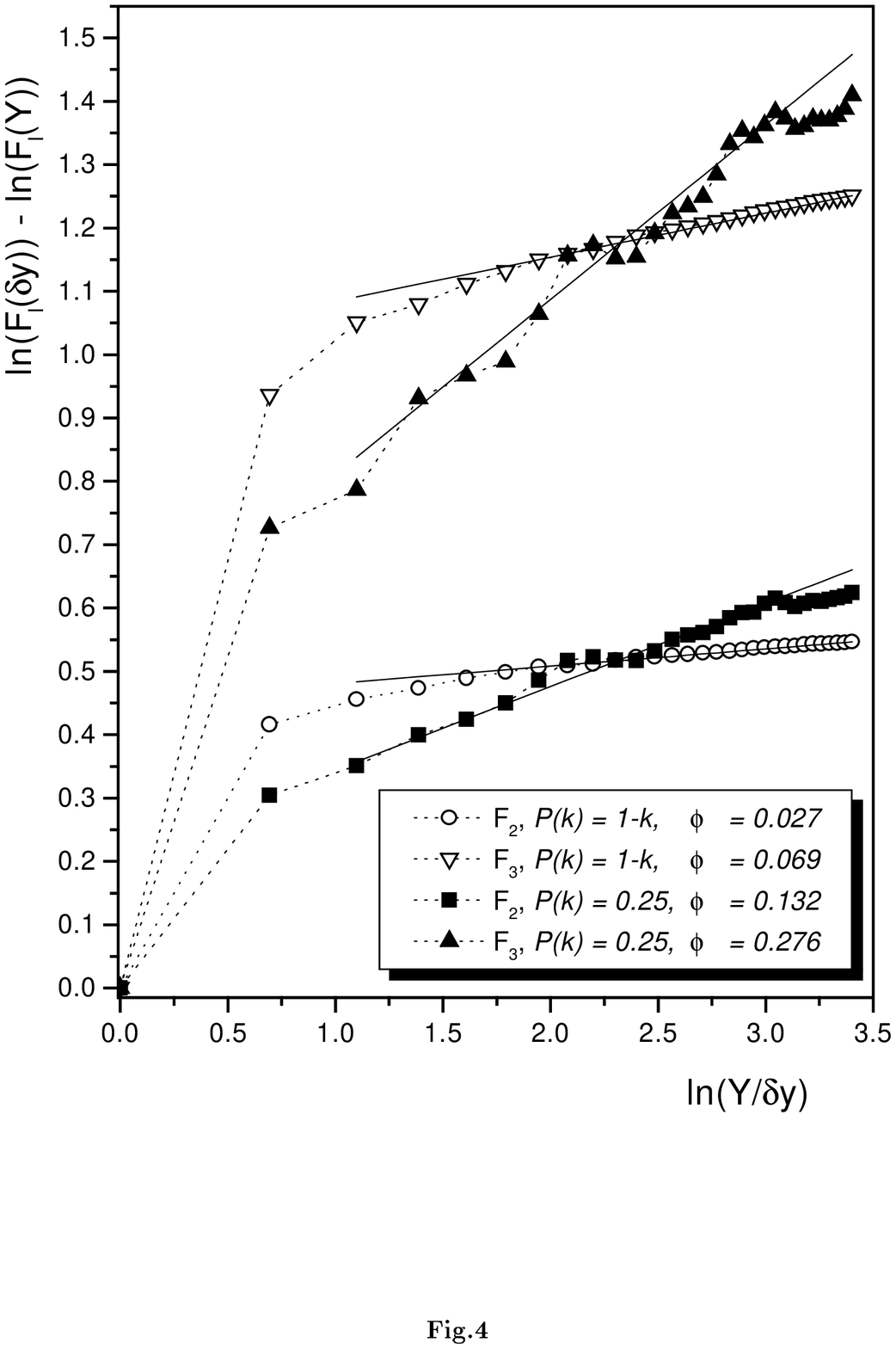}
\end{picture}
\end{figure}

\newpage
\begin{figure}[h]
\setlength{\unitlength}{1cm}
\begin{picture}(25.,16.5)
\includegraphics{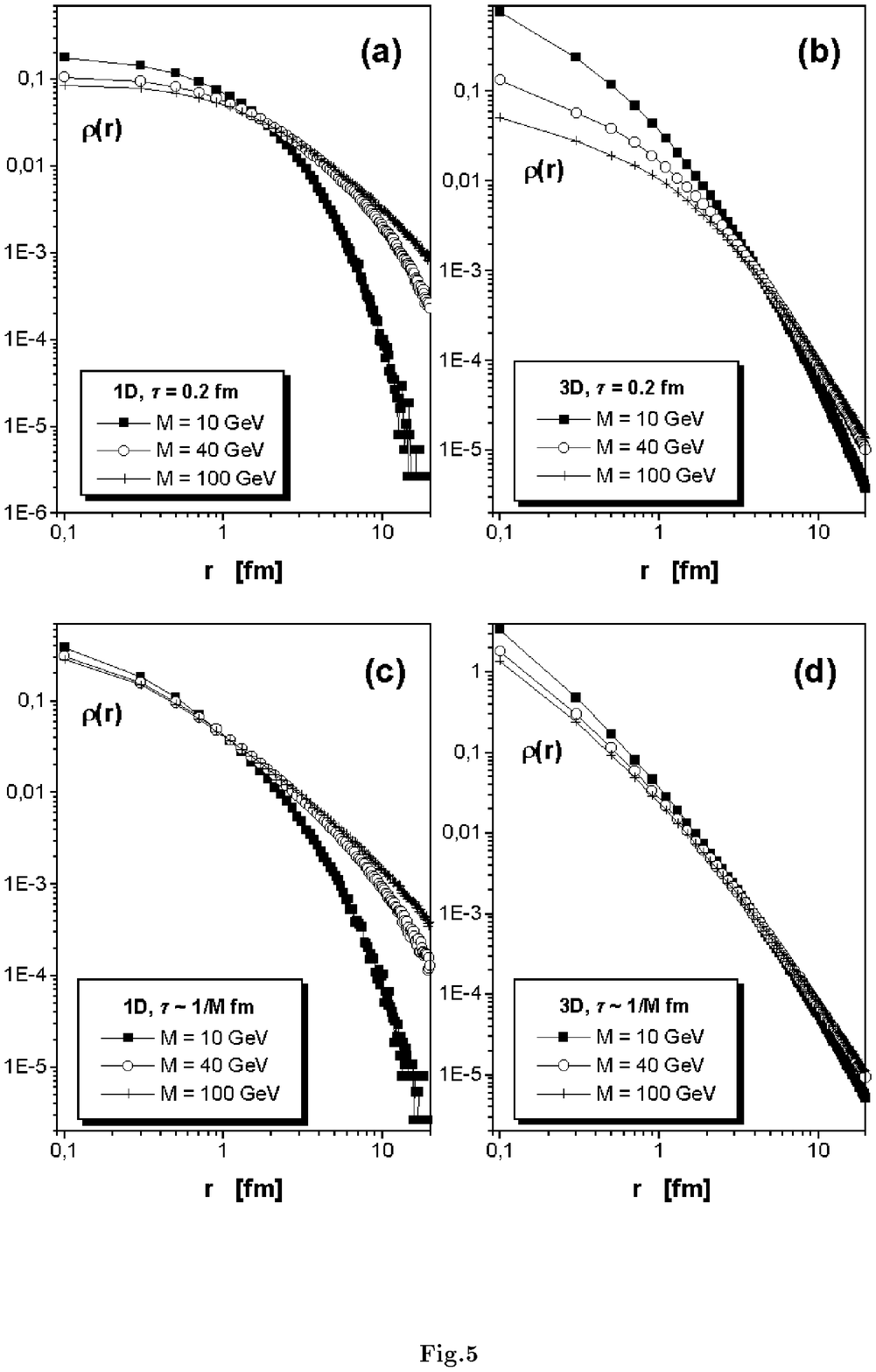}
\end{picture}
\end{figure}

\newpage
\begin{figure}[h]
\setlength{\unitlength}{1cm}
\begin{picture}(25.,16.5)
\includegraphics{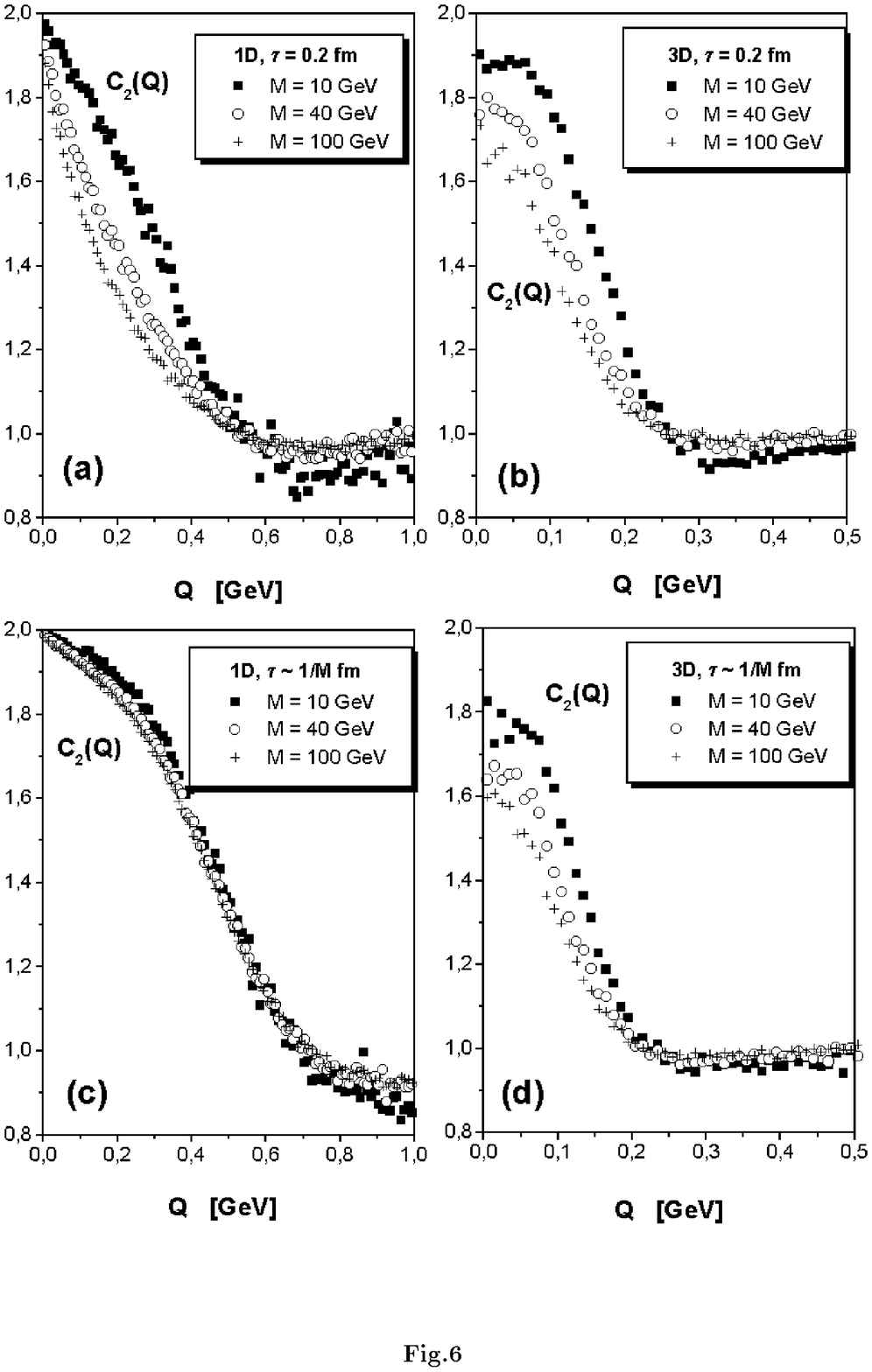}
\end{picture}
\end{figure}


\newpage


\begin{thebibliography}{99}

 \bibitem{CAS} For recent reviews see: E.A.De Wolf, I.M.Dremin and
               W.Kittel, {\sl Phys. Rep.} {\bf 270} (1996) 1;
               P.Bo\.zek, M.P\l oszajczak and R.Botet, {\sl Phys.
               Rep.} {\bf 252} (1995) 101;
               I.V.Andreev, I.M.Dremin, M.Biyajima and N.Suzuki,
               {\sl Int. J. Mod. Phys.} {\bf A10} (1995) 3951
               (and references therein).

 \bibitem{BEC} Cf. the following reviews for the most recent 
               (complementary) presentations of the Bose-Einstein 
               correlations: 
               R.M.Weiner, {\it Boson interferometry in high energy
               physics}, hep-ph/9904389, to be published in
               {\sl Phys. Rep.} (1999);
               U.A.Wiedemann and U.Heinz, {\it Particle
               interferometry for relativistic heavy-ion collisions},
               nucl-th/9901094, to be published in {\sl Phys. Rep.}
               (1999). 

 \bibitem{B} A.Bia\l as, {\sl Acta Phys. Polon.} {\bf B23} (1992)
             561; cf. also A.Bia\l as, {\sl Nucl. Phys.} {\bf A545}
             (1992) 285c and in Proc. XXVII Int. Conf. High Energy
             Phys., 20-27 July 1994, Glasgow, UK, Vol.II, p. 1287,
             eds. P.J.Bursey et al, IoP Pub., Bristol, UK. 

 \bibitem{B98} For the most recent summary of the problem cf. 
               A.Bia\l as in Proc. of XXVII Int. Symp. Multiparticle
               Dynamics, Delphi, Greece, 6-11 Sept. 1998 
               (hep-ph/9812457) and references therein.

 \bibitem{PPT} J.Pi\u{s}\'ut, N.Pi\u{s}\'utov\'a and B.Tom\'asik,
               {\sl Acta Phys. Slov.} {\bf 46} (1996) 517.

 \bibitem{SS} I.Sarcevic and H.Satz, {\sl Phys. Lett.} {\bf B233}
              (1989) 251.
              
 \bibitem{GEHW} K.Geiger, J.Ellis, U.Heinz and U.A.Wiedemann,
                {\it Bose-Einstein Correlations in a Space-Time Approach
                to $e^+e^-$ Annihilation into Hadrons}, CERN-TH/98-345
                (hep-ph/9811270).

 \bibitem{ZAL} A.Bia\l as and K.Zalewski, {\sl Phys. Lett.} {\bf
               B238} (1990) 413; A.Bia\l as, A.Szczerba and
               K.Zalewski, {\sl Z. Phys.} {\bf C46} (1990) 163; cf.
               also \cite{CAS}. 

 \bibitem{CARR} P.Carruthers, {\sl Int. J. Mod. Phys.} {\bf A4} (1989)
                5589.

 \bibitem{ID} I.M.Dremin, {\sl Pisma ZETF} {\bf 24} (1987) 505 
              [{\sl JETP Lett.} {\bf 45} (1987) 643].

 \bibitem{HYDRO} L.D.Landau and S.Z.Bilenkij, {\sl Nuovo Cim. Suppl.}
                 {\bf 3} (1956) 15; E.V.Shuryak, {\sl Phys. Rep.} {\bf
                 61} (1980) 71 or R.B.Clare and D.Strottmann, {\sl
                 Phys. Rep.} {\bf 141} (1986) 177.

 \bibitem{1DIM} M.Chaichan and H.Satz, {\sl Phys. Lett.} {\bf B50} (1974)
                362.

 \bibitem{EIWW} G.Wilk and Z.W\l odarczyk, {\sl Phys. Rev.} {\bf D43}
                (1991) 794.

 \bibitem{BL} M.Bla\v{z}ek, {\sl Int. J. Mod. Phys.} {\bf A12} (1997) 839.

 \bibitem{OMT} T.Osada, M.Maruyama and F.Takagi, {\sl Phys. Rev.} {\bf D59}
               014024. 
               First calculations demonstrating the
               bunching effect in the rapidity distributions due to
               Bose statistics are due to E.H.De Groot and H.Satz, 
               {\sl Nucl. Phys.} {\bf B130} (1977) 257.
               
 \bibitem{PIS} J.Masarik, A.Nogov\'a, J.Pi\v{s}\'ut and N.Pi\v{s}\'utova,
               {\sl Phys.} {\bf C75} (1997) 95 and {\sl Acta Phys.
               Slov.} {\bf 47} (1997) 63. Cf. also M.\v{S}iket, J.Masarik,
               A.Nogov\'a, J.Pi\v{s}\'ut and N.Pi\v{s}\'utova, {\sl
               Acta Phys. Slov.} {\bf 48} (1998) 563.
 
 \bibitem{Z} W.A.Zajc, {\sl Phys. Rev.} {\bf D35} (1987) 3396.               

 \bibitem{DL} I.M.Dremin and B.B.Levchenko, {\sl Phys. Lett.} {\bf B292}
              (1992) 155.

 \bibitem{HWA} Z.Cao and R.C.Hwa, {\sl Phys. Rev. Lett.} {\bf 75} (1995)
               1268; {\sl Phys. Rev.} {\bf D53} (1996) 6608; {\it ibidem}
               {\bf D54} (1996) 6674. Cf. also R.C.Hwa and Y.Wu, OITS
               668 preprint (hep-ph/9904213) and references therein.

 \bibitem{MB} R.Shimoda, M.Biyajima and N.Suzuki, {\sl Prog. Theor.
              Phys.} {\bf 89} (1993) 697 and T.Mizoguchi, M.Biyajima
              and T.Kagea, {\sl Prog. Thoer. Phys.} {\bf 91} (1994) 905.

 \bibitem{MV} M.Greiner, H.C.Eggers and P.Lipa, {\sl Phys. Rev. Lett.}
              {\bf 80} (1998) 5333; M.Greiner, J.Schmiegel,
              F.Eickemeyer, P.Lipa and H.C.Eggers, {\sl Phys. Rev.}
              {\bf E58} (1998) 554. For very recent studies on this
              subject see A.Bia\l as and J.Czy\.zewski, {\it Density
              correlators in a self-similar cascade}, TPJU 5/99, June
              1999; hep-ph/9906390 and references therein.
              
\end{thebibliography}
\end{document}